\documentclass[preprint,aps,tightenlines,showpacs,nofootinbib,superscriptaddress]{revtex4}
\usepackage{color}
\usepackage{enumerate}
\usepackage{latexsym}
\usepackage[dvips]{graphicx}
\def\be{\begin{equation}}
\def\ee{\end{equation}}
\def\ba{\begin{eqnarray}}
\def\ea{\end{eqnarray}}
\newcommand{\beq}{\begin{eqnarray}}
\newcommand{\eeq}{\end{eqnarray}}
\newcommand{\eq}{eqnarray}

\newcommand{\al}{{\alpha}}
\newcommand{\ci}{\cite}

\newcommand{\ep}{{\epsilon}}

\newcommand{\de}{{\delta}}

\newcommand{\La}{{\Lambda}}
\newcommand{\m}{{\mu}}
\newcommand{\n}{{\nu}}

\newcommand{\pa}{{\partial}}
\newcommand{\no}{{\nonumber}}
\newcommand{\f}{\frac}
\newcommand{\wt}{\widetilde}
\newcommand{\ra}{\rightarrow}

\newcommand{\Sch}{Schwarzschild }

\begin{document}

\preprint{arXiv:1608.00445v3 [hep-th]}

\title{On a New Approach for Constructing Wormholes in Einstein-Born-Infeld Gravity}

\author{Jin Young Kim \footnote{E-mail address: jykim@kunsan.ac.kr}}
\affiliation{Department of Physics, Kunsan National University,
Kunsan 54150, Korea}

\author{Mu-In Park \footnote{E-mail address: muinpark@gmail.com, Corresponding author}}
\affiliation{Research Institute for Basic Science, Sogang University,
Seoul, 121-742, Korea}

\begin{abstract}
We study a new approach for the wormhole construction in Einstein-Born-Infeld
gravity, which does not require exotic matters in the Einstein equation.
 The Born-Infeld field equation is not modified from {\it coordinate independent} conditions of continuous metric tensor and its derivatives, even though the Born-Infeld fields have discontinuities in their derivatives at the throat in general. We study the relation of the newly introduced conditions with the usual continuity equation for the energy-momentum tensor and the gravitational Bianchi identity. We find that there is no violation of energy conditions for the Born-Infeld fields contrary to the usual approaches.
 The exoticity of energy-momentum tensor is not essential for sustaining
wormholes. Some open problems are discussed.
\end{abstract}


\maketitle

\newpage

\section{Introduction}

Wormholes are 
non-singular space-time structures connecting two (or more) different universes or parts of the same universe \cite{Whee}. However, in the conventional approaches for (traversable) wormholes, there are problems in the ``naturalness''. First, the hypothetical exotic matters, which violate the energy conditions in the standard general relativity (GR) but are essential for sustaining the throats of wormholes, have never been discovered \cite{Elli}. Second, the usual cuts and pastes of the throats and subsequent putting of the hypothetical matters to the throats by hand to satisfy the Einstein equation \ci{Viss} are too artificial to be considered as a natural process. Moreover, we don't know much about the formation mechanism of wormholes due to the gravitationally repulsive nature of the exotic matters.

To circumvent the obstacle by exotic matters, one may consider other extended
theories of gravity so that the exotic matters are naturally generated. For example, one may consider non-minimally coupled scalar-tensor gravity theories \ci{Agne}, higher curvature gravities \ci{Hoch}, higher dimensional gravities \ci{Chod}, and brane-worlds theories \ci{Bron}. (For a more complete and modern review, see Ref. \ci{Lobo} and references therein.) On the other hand, to cure the problem of artificiality of the conventional cuts and pastes construction of throats of wormholes, a new approach for the wormhole construction has been proposed recently \ci{Cant,Kim}. In the new approach, the throat is defined as the place where the solutions are smoothly joined. There, the metric and its derivatives are continuous so that the exotic matters are not introduced at the throats.

From the new definition, throats can not be constructed arbitrarily contrary to the conventional cuts and pastes approach. For example, we consider a four dimensional spherically symmetric wormhole connecting two remote parts of the same universe (or mathematically equivalently, the reflection symmetric two universes). The metric is described by
\begin{\eq}
  ds^2=-N_{\pm}(r)^2 c^2 dt^2+\frac{dr^2}{f_{\pm}(r)}+r^2
\left(d\theta^2+\sin^2\theta d\phi^2\right)
\label{wormhole}
\end{\eq}
with two coordinate patches, each one covering the range $[r_0, +\infty)$.
Then, the radius of throat $r_0$ is defined as
\begin{\eq}
\left.\f{dN_{\pm}}{dr}\right|_{r_0}=\left.\f{df_{\pm}}{dr}\right|_{r_0} =0
\label{throat}
\end{\eq}
with the usual matching condition
\begin{\eq}
N_+(r_0)=N_-(r_0), ~f_+(r_0)=f_-(r_0)
\label{metric_contin}
\end{\eq}
by demanding that the metric and the \Sch coordinates be continuous{, though not generally smooth,} across the throat. If there exists a coordinate patch ${\cal M}_+$ in which the singularities are absent for all values of $r \geq r_0$, one can construct a smooth regular wormhole-like geometry, which may or may not be traversable depending on the location of $r_0$, by joining ${\cal M}_+$ and its mirror patch ${\cal M}_-$ at the throat $r_0$. Here, it is important to note that, in the new approach, $f_\pm(r_0)$ needs {\it not} to be vanished
in contrast to the Morris-Thorne approach \ci{Elli}, while the quantities ${dN_{\pm}(r_0)}/{dr},~ {df_{\pm}(r_0)}/{dr}$ in (\ref{throat}) need {\it not} to be vanished in both the Morris-Thorne approach \ci{Elli} and the cuts and pastes approach of Visser \ci{Viss}.

So, in the new approach, we have two options for exact solutions of the same Einstein equation when the throat point $r_0$ exists. We may consider either the usual solutions for compact objects, like the black holes which may have singularities (but usually shielded by the horizons) generally, or wormhole-like objects which do not have singularities in the whole space-time domain. We do not have a priori reason to choose only one of the two options so that we can consider both options equally. This is in contrast to the Morris-Thorne-type traversable wormholes, which can not co-exist with the solutions of black holes with the same conserved {charges}. In this context, we call the new type of wormholes as ``natural wormholes'' and the their throats as the {\it natural} throats to distinguish these from the previously known wormholes.

Originally the new wormhole solutions \ci{Cant} have been studied in
Ho\v{r}ava (or Ho\v{r}ava-Lifshitz) gravity, which has been
proposed as a renormalizable quantum gravity based on different scaling
dimensions between the space and time coordinates \ci{Hora}. But the new approach can be applied to other gravity
theories as well if the natural throat exists. However, the existence
of the throat for black holes with rotation would be very difficult
unless some coincidences occur. For the Kerr black hole in GR, there
are more metric functions and there is no solution for the throat
where all the metric functions join smoothly simultaneously. This
implies that the wormhole throat in the spherically symmetric
configurations could be easily destroyed by adding other
conserved {charges}. Conversely, the wormhole throat
could be formed only after losing all {charges} except
the mass \ci{Kim}. It has been also argued that the situation with
electromagnetic charges would be similar because there
are additional gauge fields to be joined smoothly at the same throat. One can easily show that there is no solution for the throat where the electromagnetic fields $F^{\pm}_{\mu\nu}$ join smoothly, ${dF^{\pm}_{\mu\nu}(r_0)}/{dr}=0,~ F^+_{\mu\nu} (r_0)=F^-_{\mu\nu}(r_0)$, in addition to the conditions for the metric (\ref{metric_contin}).

The purpose of this paper is to see whether the concept of natural wormholes
could be also generalized to the charged cases or not. In particular, in
order to see the generic role of charges for non-linear electromagnetic
fields at short distances, we consider the Born-Infeld (BI) action, instead
of the usual Maxwell's, without modifying the Einstein-Hilbert action at
short distances.

The organization of this paper is as follow. In Sec. II, we set up the basic
equations of Einstein-Born-Infeld gravity with a cosmological constant and
study the exact solutions for spherically symmetric black holes and their
physical properties, including the black hole thermodynamics and phase structures.
In Sec. III, we study the natural wormhole geometry as the solution of the Einstein equation without introducing additional matters at the throat based on the construction sketched in Sec. I. We show that the BI field equation without additional terms at the throat is still valid even with the throat from
some reasonable conditions of ``coordinate independent'' continuity of metric
tensors, the preservation of the usual gravitational Bianchi identity, and
the continuity equation of energy-momentum tensors for the BI fields.
In Sec. IV, we conclude with several discussions. Throughout this paper, we
use the conventional units for the speed of light $c$ and the Boltzman's
constant $k_B$, $c=k_B=1$, but keep the Newton's constant $G$ and the
Planck's constant $\hbar$ unless stated otherwise.

\section{Black hole solutions in Einstein-Born-Infeld gravity}

Born-Infeld action of non-linear electrodynamics was originally
introduced to solve the infinite self-energy of a point charge in Maxwell's linear electrodynamics \cite{Born}.
When this action is combined with Einstein action to study the gravity
of charged objects, the short-distance behavior of the metric is modified \cite{Hoff}. With the recent development of D-branes, the Born-Infeld-type action has attracted renewed interests as an effective action for
low-energy superstring theory \ci{Leig}. There have been extensive studies
about black hole solutions in the Einstein gravity coupled to Born-Infeld
electrodynamics (Einstein-Born-Infeld (EBI) gravity)
\cite{Garc,Demi,Oliv,Rash,Ferd,Dey,Cai,Myun,Guna,Zou,Fern}. In this section, we briefly describe EBI gravity and its black hole solutions.

Since our results do not depend much on the space-time dimensions $D \geq 4$,
we consider the EBI gravity action with a cosmological constant $\Lambda$ in $D=3+1$ dimensions for simplicity,
\be
 S = \int d^{4} x \sqrt{-g} \left [ \frac{(R - 2\Lambda)}{16 \pi G}
  + L(F)                            \right ],
 \label{EBI}
\ee
where $L(F)$ is the BI Lagrangian density given by
 \be
 L(F) = 4 \beta^2 \left( 1 - \sqrt{ 1 + \frac{ F_{\mu\nu} F^{\mu\nu} }{2 \beta^2} }
                  \right) .
 \label{BI}
 \ee
Here, the parameter {$\beta~$} is a coupling constant with dimensions
$[{\rm length}]^{-2}$ which needs to flow to infinity to recover the
usual Maxwell electrodynamics with
$L(F)=-F_{\mu\nu} F^{\mu\nu} \equiv -{F}^2$ at low energies. The
correction terms from a finite $\beta$ represent the effect of the
non-linear BI fields. This situation is similar to the Lagrangian for a
relativistic free particle $mc^2 (1-\sqrt{1-{v}^2/c^2})$ which reduces
to the Newtonian Lagrangian $\f{1}{2} m v^2$ for $c \ra \infty$ limit.
As the speed of light $c$ gives
the upper limit for the speed $v$ of a particle, the parameter
$\beta$
gives the upper limit for the strength of fields $\sqrt{-\f{1}{2}
{F}^2}$ if ${F}^2 <0$.

\begin{figure}
\includegraphics[width=7cm,keepaspectratio]{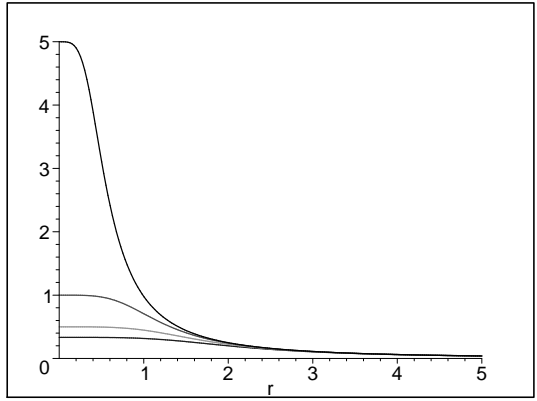}
\caption{The plots of {$E(r)$} for varying $\beta$ with a fixed $Q$. In particular, we consider $\beta=5,~1,~1/2,~1/3,$ (top to bottom) with $Q=1$.}
\label{fig:E}
\end{figure}

Taking $16 \pi G = 1$ for simplicity, the equations of motion are
obtained as
\be
\nabla_{\mu} \left ( \frac{F^{\mu\nu}}{\sqrt{ 1 + \frac{ F^2 }{2 \beta^2} }}
             \right ) = 0 ,
\label{eomforBI}
\ee
\be
 R_{\mu\nu} - \frac{1}{2}R g_{\mu\nu} + \Lambda g_{\mu \nu}=\f{1}{2} T_{\mu\nu},
 \label{eomforgrav}
 \ee
where the energy momentum tensor for BI fields is given by
 \be
  T_{\mu\nu}=g_{\mu\nu} L(F) +\frac{4 F_{\rho\mu} F^\rho_{\nu} }{\sqrt{ 1 + \frac{ F^2 }{2 \beta^2} }} .
   \label{Tmunu}
   \ee
Let us now consider a static and spherically symmetric solution with
the metric ansatz
 \be
 ds^2 = - N^2 (r) dt^2 + \f{1}{f(r)} dr^2 + r^2 (d \theta^2 + \sin^2 \theta d \phi^2).
 \ee
For the static \footnote{Here, the static property is the result of the BI field equation (\ref{eomforBI}). If we relax this assumption, we need to consider the magnetic components of the field strength as well.}
electrically charged case where the only non-vanishing component of the field strength tensor is {$F_{rt} \equiv E $}, one can find
\beq
N^2(r)=f(r).
\label{N=f}
\eeq
Then, the general solution for the BI electric field is obtained from the equation (\ref{eomforBI}) as
 \be
 E(r) = \frac{Q}{\sqrt{ r^{4} + \frac{Q^2}{\beta^2}    } } ,
 \label{Esol}
 \ee
where $Q$ is the integration constant that represents the electric
charge localized at the origin. Note that the electric field is finite
in the limit $r \to 0$, $E(r) \approx \pm |\beta|$ for $|\beta|<\infty$
due to the non-linear effect of BI fields while it reduces to the usual Coulomb field $E \approx Q/r^2$ in the limit $|\beta| \to \infty$ (Fig. \ref{fig:E}).

Now, from (\ref{N=f}) and (\ref{Esol}), the Einstein equation (\ref{eomforgrav}) reduces to
 \be
 \frac{d}{dr} (r f ) =  1 - \Lambda r^2
 + 2 \beta^2 \left ( r^2 - \sqrt{ r^{4} + \frac{ Q^2} { \beta^2}  } \right ),
 \label{f:Eq}
 \ee
and a simple integration over $r$ gives
 \be
 f = 1 -\f{2 C}{r}- \frac{\Lambda}{3} r^2
 + \f{2 \beta^2}{r} \int_0^r r^{2} \left ( 1 - \sqrt{ 1 + \frac{Q^2}{\beta^2 r^{4}} } \right ) dr,
 \label{f_int}
 \ee
where $C$ is an integration constant.

With a straightforward integration, one can express the solution in
a compact form in terms of the incomplete elliptic integral of the
first kind,
\ba
 f = 1 - \frac{2C}{r} - \frac{\Lambda}{3} r^2
 + \frac{2\beta^2}{3} r^2 \left ( 1 - \sqrt{ 1+ \frac{Q^2} { \beta^2 r^{4} } } \right )
  - \frac{4 }{3 r} \sqrt{-i {\beta Q^3}}~ {\rm EllipticF} \left(r \sqrt{\frac{i \beta} {Q}},i \right ).    \label{f:sol1}
\ea
For small $r$, this may be expanded as
 \be
 f(r) = 1 - 2 \beta Q - \frac{2C}{r} + \frac{1}{3} (2 \beta^2 - \Lambda) r^2
 -\frac{\beta^3}{5 Q}r^4+{\cal O}(r^6),
 \label{f:smaller}
 \ee
which shows a mass-like term $-2C/r$. The parameter $C$ is called an intrinsic mass \ci{Demi} and represents the mass-like behavior near the origin. However, it is important to note that this is not the usual ADM mass defined at large $r$. In order to see this, it is convenient to reorganize the integral in (\ref{f_int}) as $\int^r_0=\int^\infty_0+\int^r_\infty$ so that the integral $\int^r_\infty$ is convergent in the limit $r\ra \infty$. Then, we have another expression for the solution
\be
 f(r) = 1 - \frac{2 M}{r} - \frac{\Lambda}{3} r^2
 + \frac{2 }{3} \beta^2 r^2 \left ( 1 - \sqrt{ 1+ \frac{Q^2} { \beta^2 r^{4} } } \right )
  + \frac{4}{3} \frac{Q^2} {r^{2}}~ {_2 F}_1 \left ( \frac{1}{2} , \frac{1}{4}; \frac{5}{4};
  \frac{-Q^2}{\beta^2 r^{4}} \right ),
   \label{f:sol2}
\ee
in terms of the hypergeometric function \ci{Oliv,Rash,Ferd,Dey,Guna}. Here, $M$ is another mass parameter, defined by
\beq
M &\equiv &C+ M_{0}, \\
M_{0} &=& \frac{2}{3} \sqrt{ \frac{\beta Q^3}{\pi} } \Gamma\left( \frac{1}{4} \right) \Gamma\left( \frac{5}{4} \right),
 \eeq
where $M_{0}$ comes from the $\int^\infty_0$ part of the integral. One can find that the total mass $M$, which is {the} sum of the intrinsic mass $C$ and the (finite) self-energy of a point charge $M_0$, is the usual ADM mass
defined by the asymptotic behavior of the metric at large $r$
 \be
 f(r) = 1 - \frac{\Lambda}{3} r^2 - \frac{2 M }{r}
 +  \frac{Q^2} {r^{2}} -\f{Q^4}{20 \beta^2 r^6} + {\cal O}(r^{-10}).
 \label{f:larger}
 \ee
This shows that the physical parameters, like the mass and {the} cosmological constant, are shifted at the short distances due to non-linear effects of BI fields: The fourth term on the right hand side of (\ref{f:smaller}) represents the space with an effective cosmological constant $\La_{\rm eff}=\La-2 \beta^2$, which {can} behave like an anti de-Sitter space $(\La_{\rm eff}<0)$ near the origin even though it behaves like a {flat} $(\La=0)$ or a {de-Sitter} $(\La>0)$ space at the asymptotic infinity.

The solution (\ref{f:sol1}) or (\ref{f:sol2}) has a curvature singularity at the origin, $r=0$,
\beq
R&=&\f{4 \beta Q}{r^2} +4 (\La-2 \beta^2) + \f{6 \beta^3}{Q}r^2+ {\cal O}(r^4), \no \\
R^{\m\n}R_{\m\n}&=&\f{8 \beta^2 Q^2}{r^4} +\f{8 \beta Q (\La-2 \beta^2)}{r^2}+
4 (\La^2-4 \La \beta^2+6 \beta^4) +{\cal O}(r^2), \no \\
R^{\m\n\al\beta}R_{\m\n\al\beta}&=&\f{48 C^2}{r^6}
+\f{32 \beta Q C}{r^5} +\f{16 \beta^2 Q^2}{r^4} +{\cal O}(r^{-2}).
\label{sing}
\eeq
Note that the leading singularity near $r=0$ is of the order
${\cal O}(r^{-6})$ in the Riemann tensor square, which is the same as
that of \Sch (Sch) black hole. But this singularity is absent for the
marginal case of $C=0$ ($M=M_{0}$) due to the regular behavior of the
solution near the origin (\ref{f:smaller}), which is the same as that of Reissner-Nodstrom (RN) black hole \footnote{One can not directly obtain the singular behavior of RN black hole near $r=0$ from (\ref{sing}) as one can not obtain RN black hole solution from (\ref{f:smaller}) by taking $|\beta| \ra \infty$ limit because $|\beta| <\infty$ has been implicitly assumed in the expansions of (\ref{f:smaller}) and (\ref{sing}). One can obtain RN results only from (\ref{f:larger}) which is valid always unless $\beta = 0$.}. On the other hand, it can be shown that there is no curvature singularity at the horizon, which is defined as the solution for $f(r)=0$ in our case, as expected \ci{Oliv}.

The solution can have two horizons generally and the Hawking temperature for the outer horizon $r_+$ is given by
\beq
T_H &\equiv& \f{\hbar}{4 \pi} \left. \f{df}{dr} \right|_{r=r_+} \no \\
&=&\f{\hbar}{4 \pi} \left[ \f{1}{r_+}-\La r_+ +2 \beta^2 r_+ \left(1-\sqrt{1+\f{Q^2}{\beta r_+^4}} \right) \right].
\eeq
There exists an extremal black hole limit of vanishing temperature, where the outer horizon $r_+$ meets the inner horizon $r_-$ at
\beq
r^*_+=\sqrt{
\f{\La-2 \beta^2
\pm 2 \sqrt{(\beta^2-\La/2)^2+\La(\La-4 \beta^2)(\beta^2 Q^2-1/4)}
}
{\La(\La-4 \beta^2)}
}
\label{r:ext}
\eeq
for a non-vanishing cosmological constant $\La$ and
\beq
r^*_+=\sqrt{Q^2-
\f{1}{4 \beta^2}}
\eeq
for vanishing $\La$. At the extremal point, the ADM mass $M$
\beq
M=\f{r_+}{2} \left[ 1 -  \frac{\Lambda}{3} r_+^2
 + \frac{2 }{3} \beta^2 r_+^2 \left ( 1 - \sqrt{ 1+ \frac{Q^2} { \beta^2 r_+^{4} } } \right )
  + \frac{4}{3} \frac{Q^2} {r_+^{2}}~ {_2 F}_1 \left ( \frac{1}{2} , \frac{1}{4}; \frac{5}{4};
  \frac{-Q^2}{\beta^2 r_+^{4}} \right ) \right]
\eeq
gets the minimum
\beq
M^*=\f{r_+^*}{3} \left[ 1
  + \frac{2 Q^2} {{r_+^*}^{2}}~ {_2 F}_1 \left ( \frac{1}{2} , \frac{1}{4}; \frac{5}{4};
  \frac{-Q^2}{\beta^2 {r_+^*}^{4}} \right ) \right].
\eeq

The first law of black hole thermodynamics is found as
\beq
dM=T_+ dS_{\rm BH} +A_0(r_+) dQ,
\eeq
with the usual Bekenstein-Hawking entropy formula by
\beq
S_{\rm BH}=\f{\pi r_+^2}{\hbar G},
\eeq
and the scalar potential {\ci{Rash}}
\beq
A_0(r)=\frac{1}{r}~ {_2 F}_1 \left ( \frac{1}{2} , \frac{1}{4}; \frac{5}{4};
  \frac{-Q^2}{\beta^2 {r}^{4}} \right ).
\eeq

Now, from the result (\ref{r:ext}), one can classify the black holes in terms of the values of $\beta Q$ and the cosmological constant $\La$ . \\

\begin{figure}
\includegraphics[width=7cm,keepaspectratio]{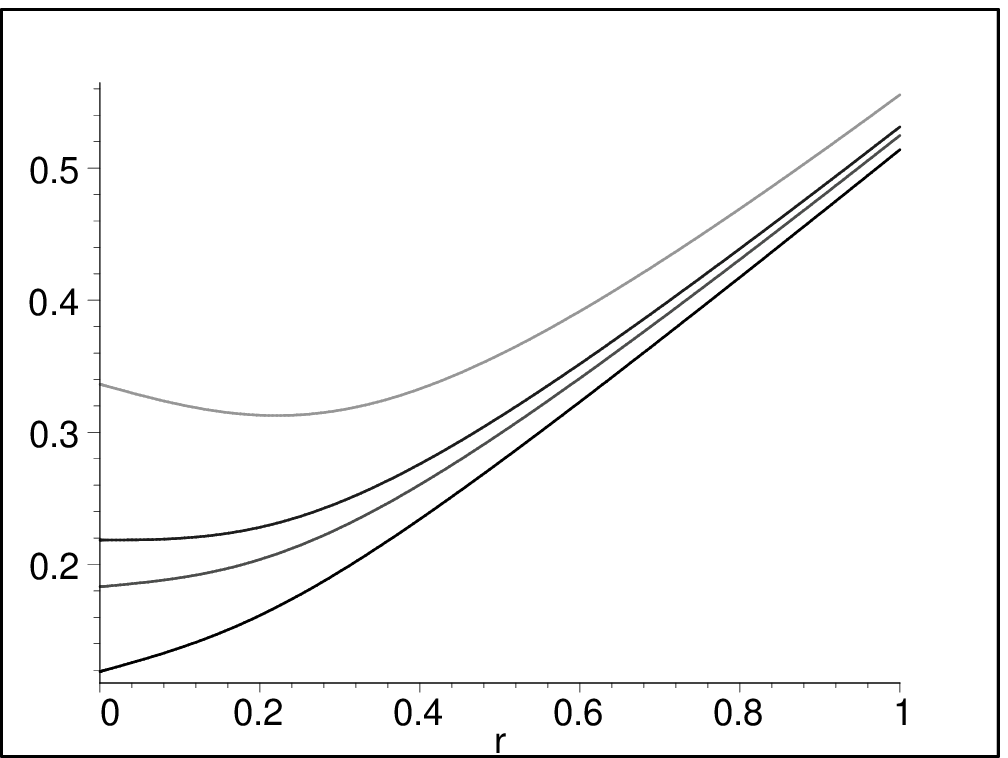}
\qquad
\includegraphics[width=7cm,keepaspectratio]{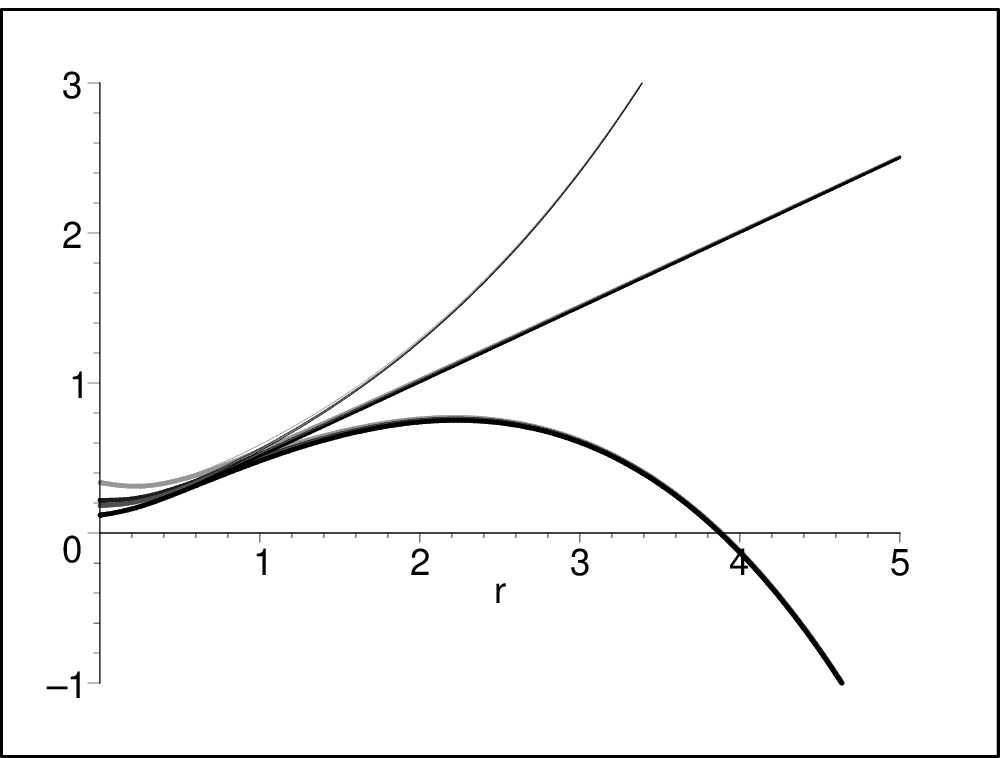}
\caption{The plots of the ADM mass $M$ vs. the black hole horizon radius $r_+$ for varying $\beta Q$ with a fixed cosmological constant $\Lambda$. For small $r_+$, there is no significant difference for different values of cosmological constant (left). The marginal mass $M_0$ is given by the mass value at $r_+=0$. The top two curves represent $M_0>M^*,~M_0=M^*$ for $\beta Q>1/2,~\beta Q=1/2$, respectively, with the extremal mass $M^*$, whereas the bottom two curves represent the cases where $M^*$ is absent for $\beta Q<1/2$. We consider $\beta Q={2/3,~1/2,~2/4.5,~1/3}$ (top to bottom) with $\beta=2$. The effect of cosmological constant is important for large $r_+$ (right): ($\Lambda<0$ (thin curve), $\Lambda=0$ (medium curve), $\Lambda>0$ (thick curve)). We consider $\Lambda=\pm 1/5$ for the (A)dS cases.}
\label{fig:M}
\end{figure}

(i) $\beta Q >1/2$: In this case, the black hole may have one non-degenerate horizon (type I), {or} two non-degenerate (non-extremal) horizons or
one degenerate (extremal) horizons (type II){,} depending on the mass
for the asymptotically AdS ($\La<0$) or flat ($\La=0$), {\it i.e.},
the type I for $M \geq M_0$ and the type II for $M < M_0$ (Fig. \ref{fig:f,bQ>1/2}a, \ref{fig:f,bQ>1/2}b). There can be a phase transition from \Sch(Sch)-like type I black hole to Reissner-Nodstrom(RN)-like type II black hole when the mass becomes smaller than the marginal mass
$M_0$, which is the mass value at $r_+=0$ in Fig. \ref{fig:M}. When the mass
is smaller than the extremal mass $M^*$, which is the mass value at the
extremal point in Fig. \ref{fig:M} (top curve), the singularity at $r=0$
becomes naked as usual. In the limit $\beta \ra \infty$, only the RN-like
type II black holes are possible as in GR. On the other hand, for
asymptotically dS ($\La>0$) case, the situation is more complicated
due to the cosmological horizon. In this case, there can exist maximally
three horizons (two smaller ones for black holes and the largest one for
the cosmological horizon), depending on the mass and the cosmological
constant (Fig. \ref{fig:f,bQ>1/2}c). \\

\begin{figure}
\includegraphics[width=4.8cm,keepaspectratio]{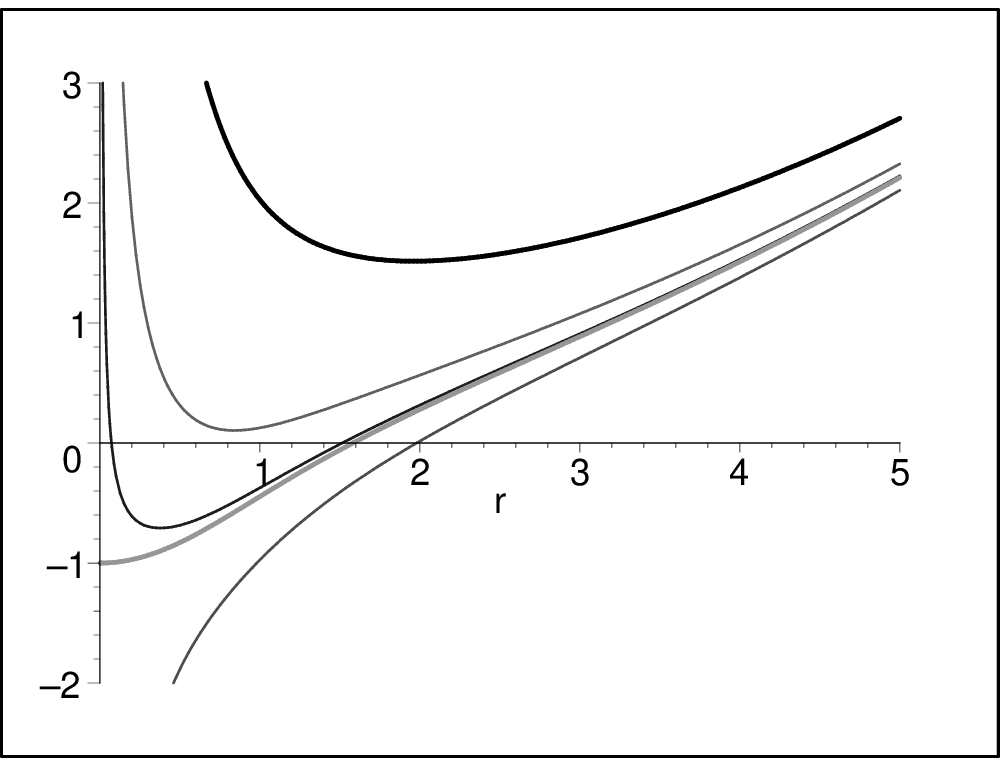}
\qquad
\includegraphics[width=4.8cm,keepaspectratio]{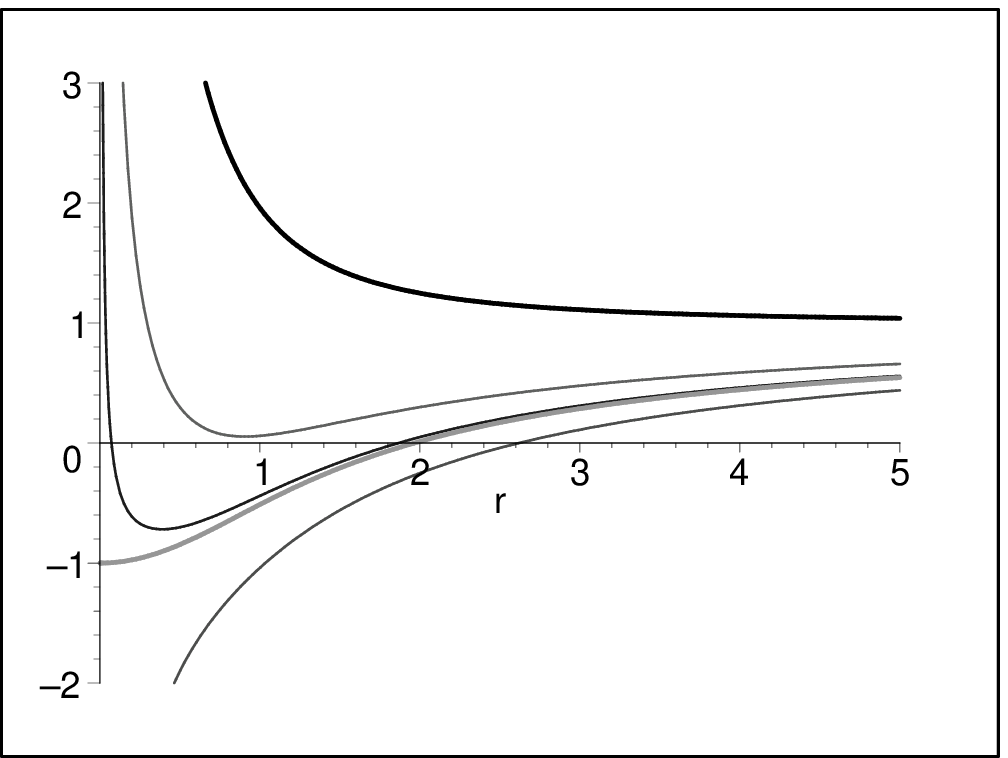}
\qquad
\includegraphics[width=4.8cm,keepaspectratio]{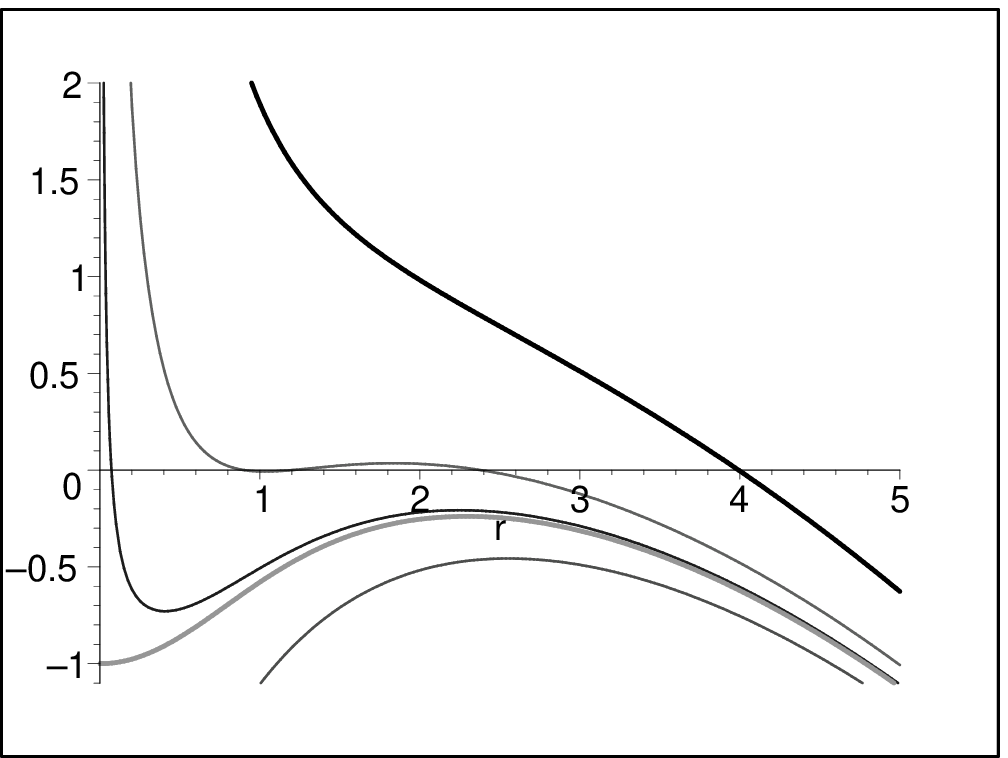}
\caption{The plots of $f(r)$ for varying $M$ with a fixed $\beta Q >1/2$ and cosmological constant $\Lambda$ ($\Lambda<0$ (left), $\Lambda=0$ (center), $\Lambda>0$ (right)). We consider $M=0,~0.95,~1.2,~M_0,~1.5$ (top to bottom) with $Q=1,\beta=1$, $M_0=1.236...$, and $\Lambda=\pm 1/5$ for the (A)dS cases.}
\label{fig:f,bQ>1/2}
\end{figure}

(ii) $\beta Q =1/2$: In this case, only the Sch-like type I black holes are
possible (Fig. \ref{fig:f,bQ=1/2}). It is peculiar that the horizon shrinks
to zero size for the marginal case $M=M_0$, even though its mass has still a non-vanishing value (Fig. \ref{fig:M}). This does not mean that the metric is a flat Minkowskian. This {rather} means that all the gravitational energy is stored in the BI electric field as a self-energy. When the mass is smaller than the marginal mass $M_0$, the singularity at $r=0$ becomes naked. Note that this is the case where the GR limit $\beta \ra \infty$ does not exist.\\

 \begin{figure}
\includegraphics[width=4.8cm,keepaspectratio]{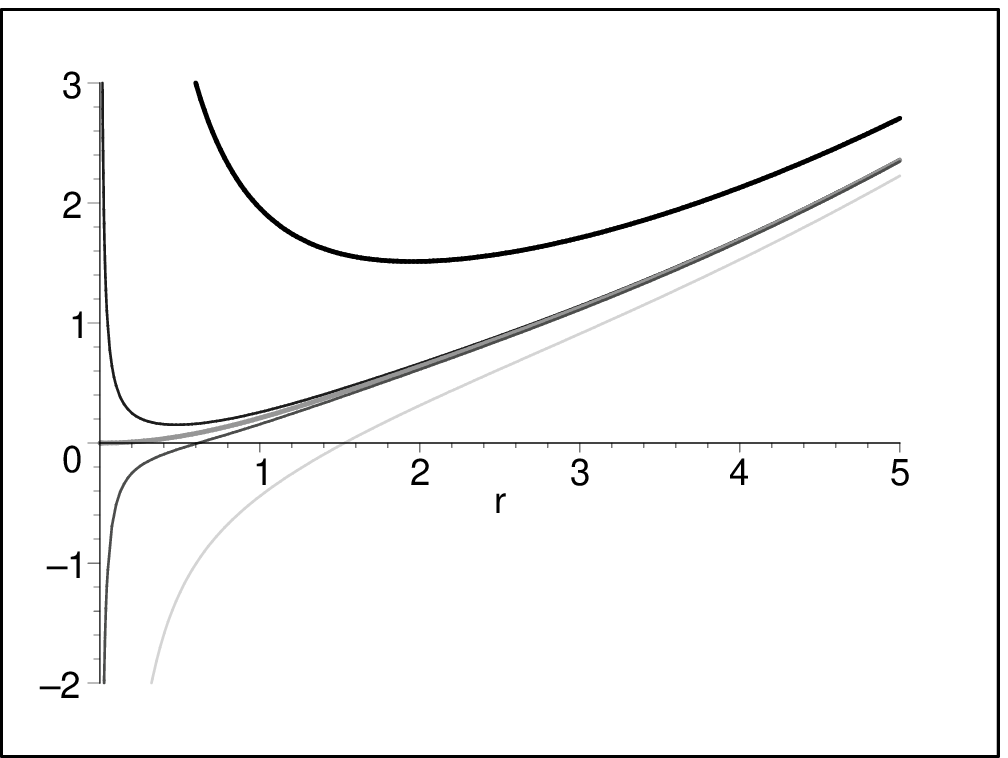}
\qquad
\includegraphics[width=4.8cm,keepaspectratio]{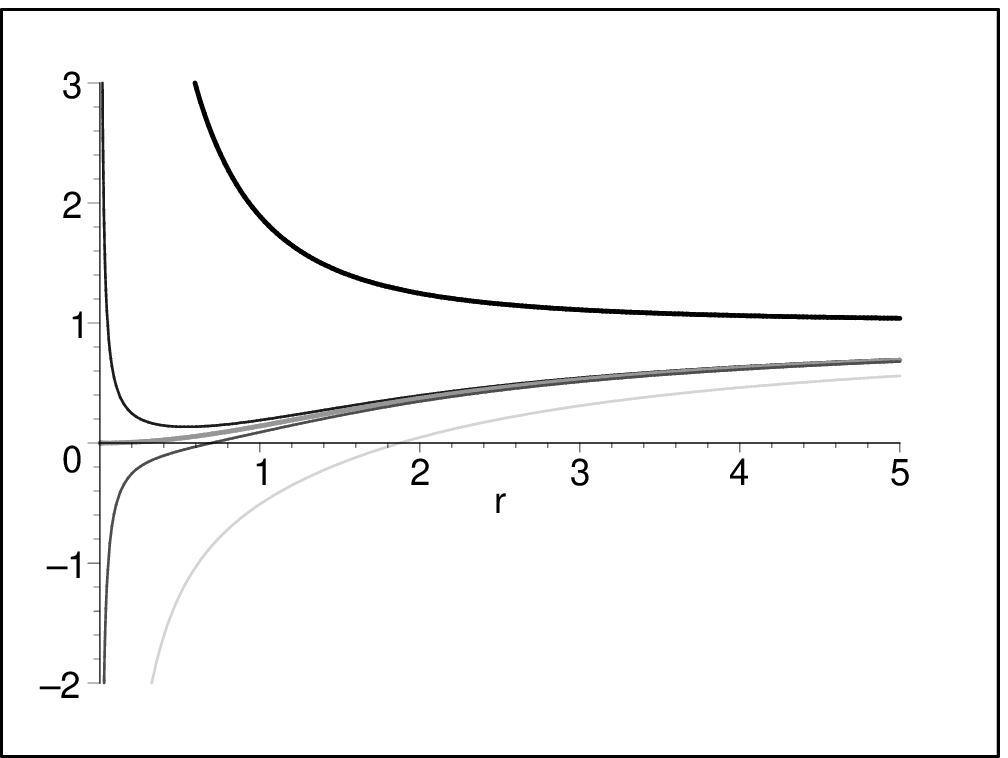}
\qquad
\includegraphics[width=4.8cm,keepaspectratio]{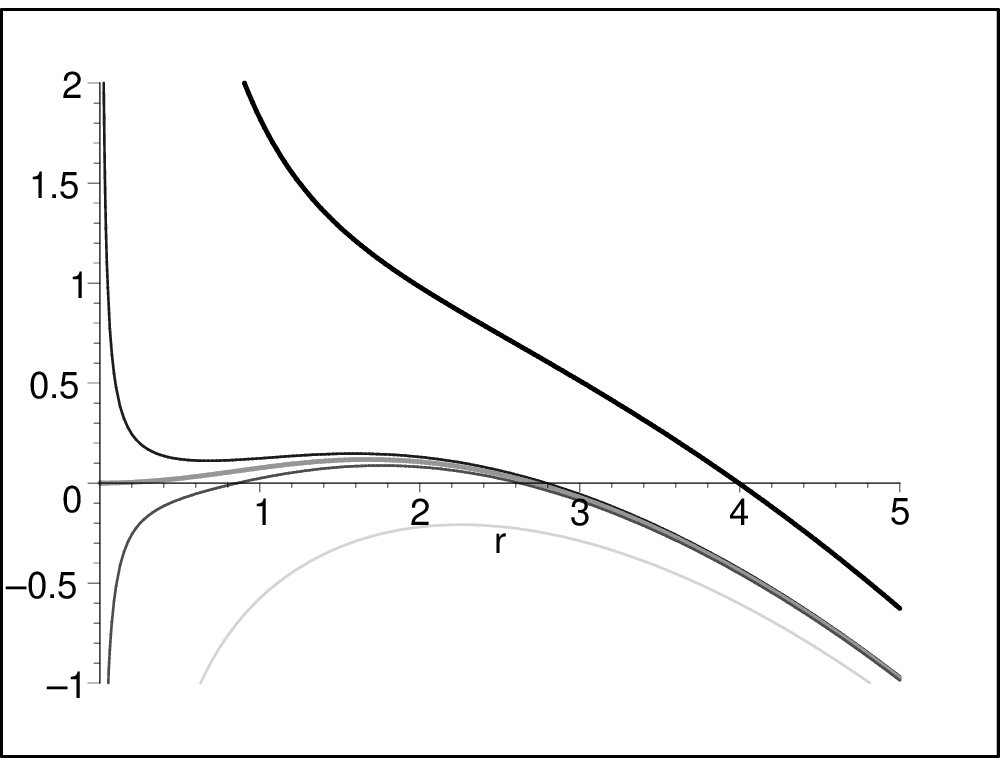}
\caption{The plots of $f(r)$ for varying $M$ with a fixed $\beta Q=1/2$ and cosmological constant $\Lambda$ ($\Lambda<0$ (left), $\Lambda=0$ (center), $\Lambda>0$ (right)).
We consider $M=0,~0.85,~M_0,~0.9,~1.2$ (top to bottom) with $Q=1,\beta=1/2$, $M_0=0.874...$, and $\Lambda=\pm 1/5$ for the (A)dS cases.}
\label{fig:f,bQ=1/2}
\end{figure}

(iii) $\beta Q <1/2$: This case is similar to the case (ii), except that the
marginal case has no (even a point) horizon so that its singularity is
naked always (Fig. \ref{fig:f,bQ<1/2}).

\begin{figure}
\includegraphics[width=4.8cm,keepaspectratio]{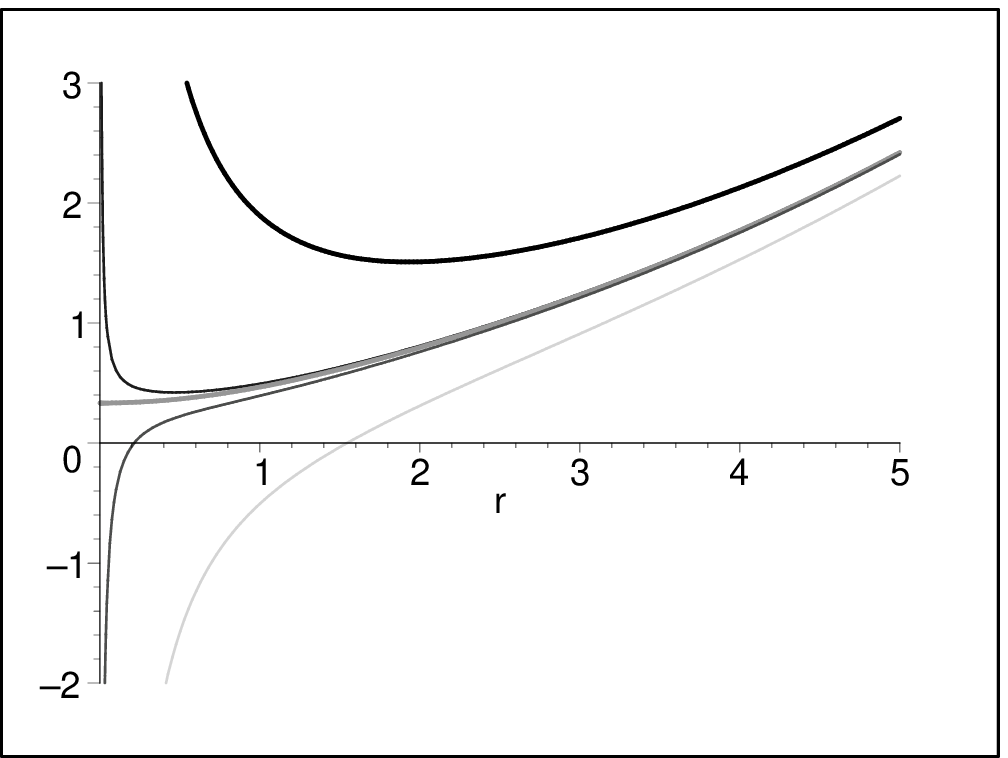}
\qquad
\includegraphics[width=4.8cm,keepaspectratio]{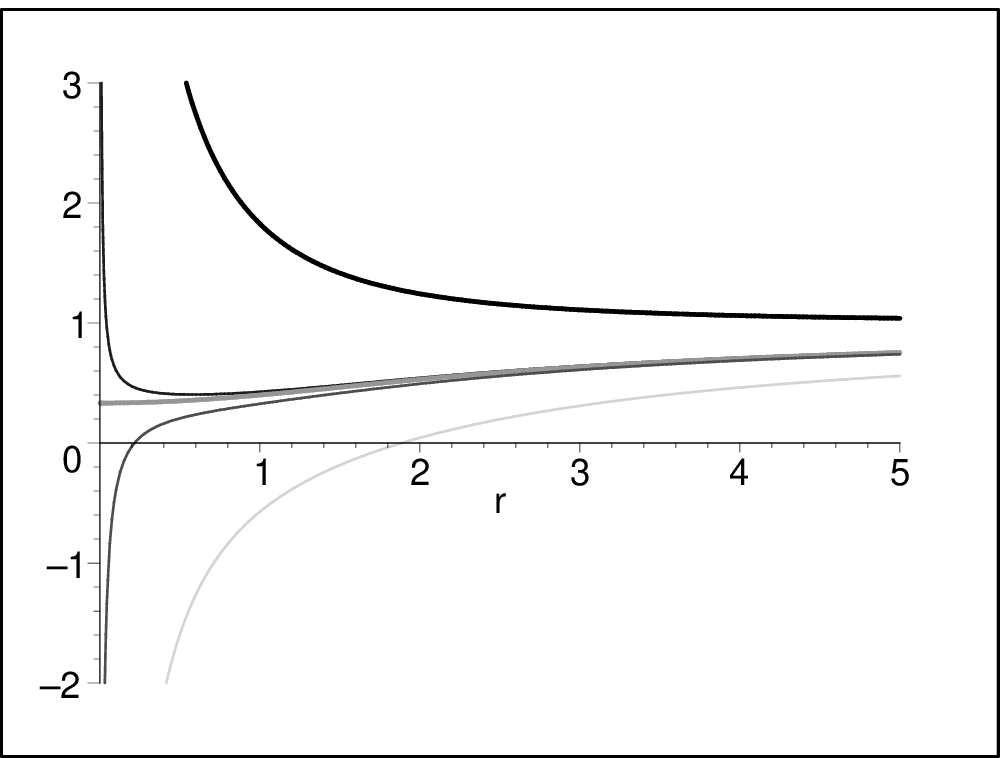}
\qquad
\includegraphics[width=4.8cm,keepaspectratio]{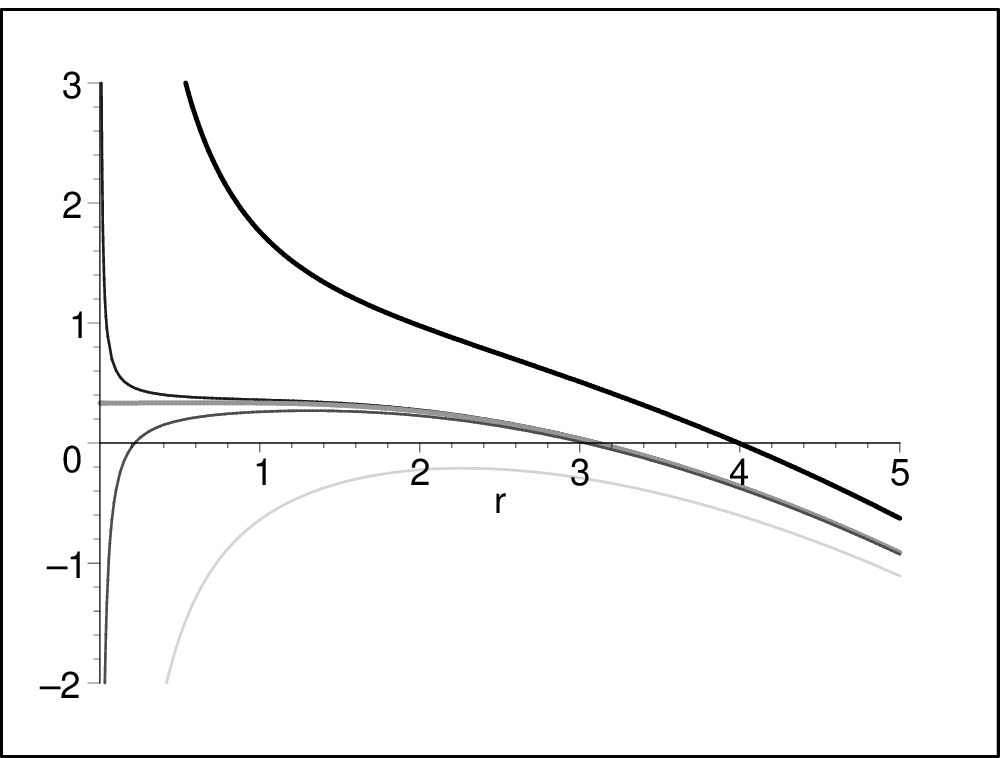}
\caption{The plots of $f(r)$ for varying $M$ with a fixed $\beta Q<1/2$ and cosmological constant $\Lambda$ ($\Lambda<0$ (left), $\Lambda=0$ (center), $\Lambda>0$ (right)).
We consider $M=0,~0.7,~M_0,~0.75,~1.2$ (top to bottom) with $Q=1,\beta=1/3$, $M_0=0.714...$, and $\Lambda=\pm 1/5$ for the (A)dS cases.}
\label{fig:f,bQ<1/2}
\end{figure}

\section{Construction of {\it natural} wormhole solutions}

The strategy to {construct} a natural wormhole is to find the throat radius 
$r_0$ defined by (\ref{throat}) with the matching condition 
(\ref{metric_contin}) \ci{Cant,Kim}. {Based on this definition},
cases (i) and (ii) in Fig. \ref{fig:f,bQ>1/2} and \ref{fig:f,bQ=1/2},
respectively, and the case (iii) in Fig. \ref{fig:f,bQ<1/2}a (AdS case) show
{the throats} depending on the mass $M$ for given values of $\beta Q$ and $\La$. {More details are as follows.}

\begin{figure}
\includegraphics[width=4.8cm,keepaspectratio]{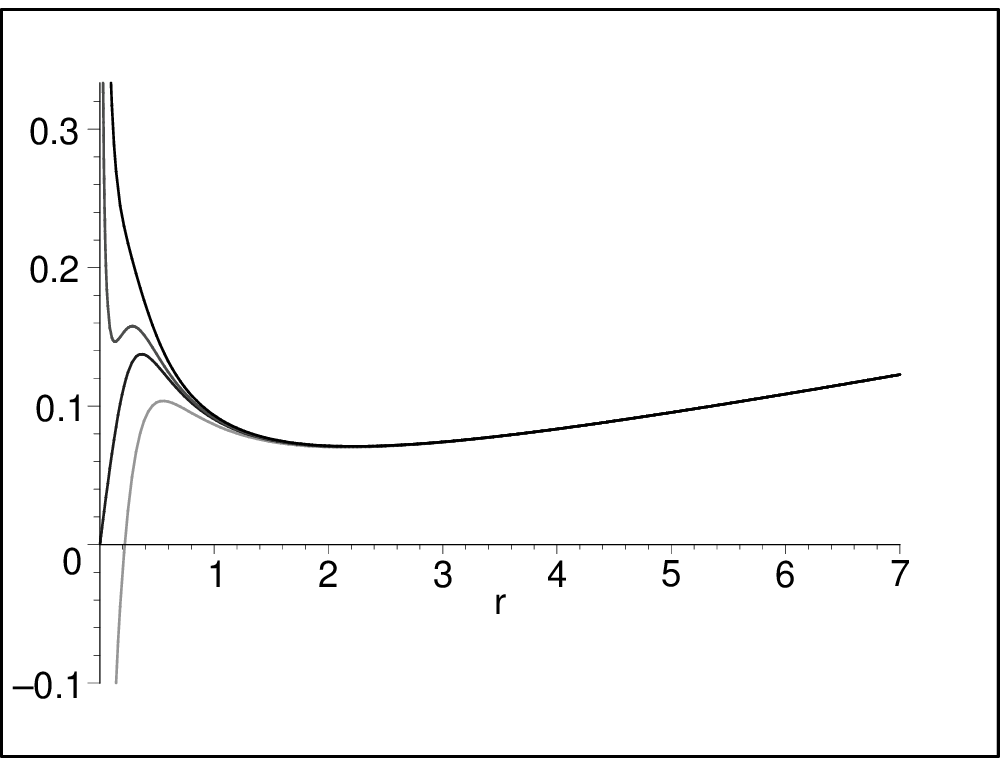}
\qquad
\includegraphics[width=4.8cm,keepaspectratio]{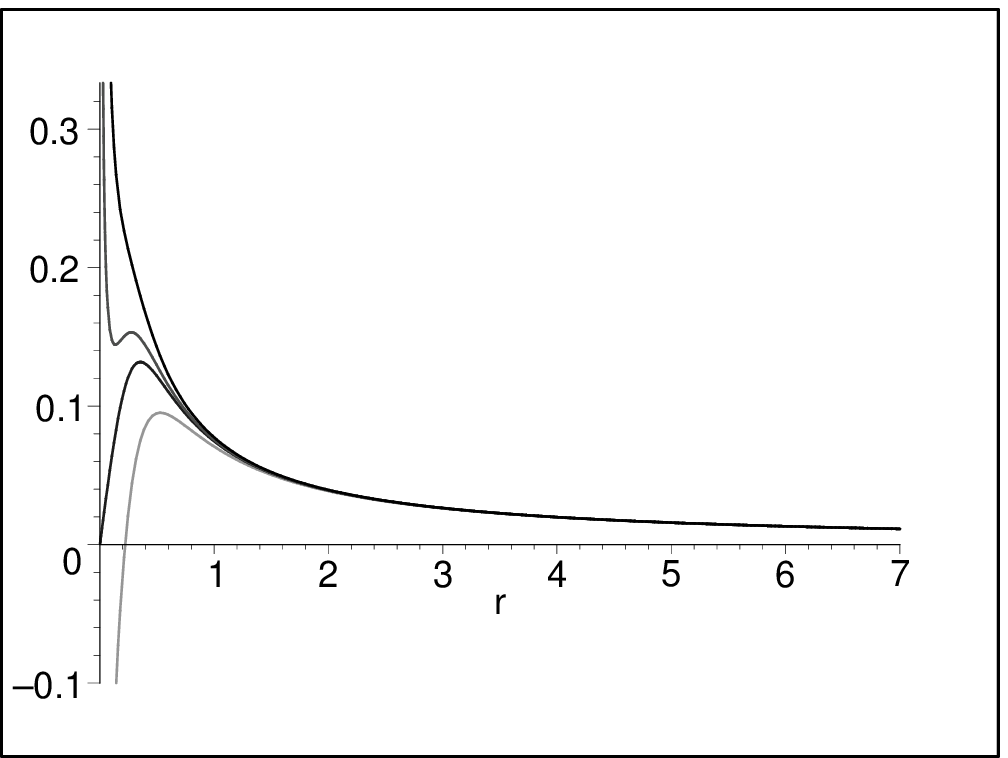}
\qquad
\includegraphics[width=4.8cm,keepaspectratio]{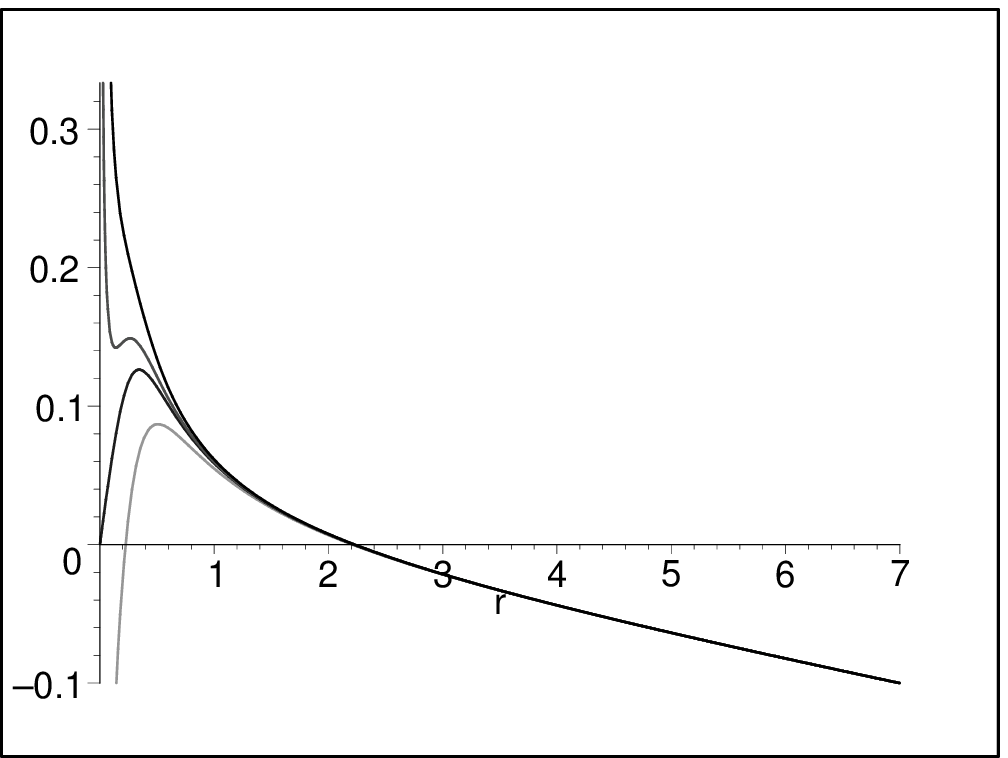}
\caption{The plots of the Hawking temperature $T_H$ vs. the black hole horizon radius $r_+$ for varying $\beta Q$ with a fixed cosmological constant $\Lambda$. We consider $\beta Q=1/3,~2/4.5,~1/2,~2/3$ (top to bottom curves) with $\beta=2$, $\Lambda=-1/5$ (left: AdS case), $\Lambda=0$ (center: flat case), $\Lambda=1/5$ (right: dS case).}
\label{fig:T}
\end{figure}

For the case (i) of $\beta Q>1/2$, there is no throat for the Sch-like type I
black holes with $M>M_0$ down to the marginal case of $M=M_0$, where a
point-like throat is generated at the origin. The zero-size ($r_0=0$) throat
can grow as the black hole loses its mass. But, down to the extremal case
of $M=M^*~(<M_0)$, where the inner and outer horizons meet at $r_+^*$ and Hawking temperature vanishes (Fig. \ref{fig:T}), the throat is inside the
outer horizon and {\it non-traversable}.\footnote{{
In the dS case
(Fig. \ref{fig:f,bQ>1/2}c), regardless of the existence of the inside throat,
there is another outside radius which satisfies the throat condition
(\ref{throat}). But this can not be considered as the wormhole throat because it does not meet the requirement that the throat occupies the
minimum radial coordinate, which has been assumed implicitly in the
definition of Sec. I.}
In principle, it could also be possible
to have a throat at the {\it maximum} radius so that one universe is smoothly
connected to another by just traveling toward the cosmological horizon.
 But, though interesting, this is not our main concern and we will not
 consider this possibility in this paper.}
As the black hole loses its mass further, there emerges and
grows a traversable throat from the extremal black hole horizon.\footnote{For the discussion about a dynamical generation of this unusual process, which is beyond the usual linear perturbation analysis, see Ref. \ci{Kim} and references therein. And this process is different from the dynamical wormhole picture suggested by Hayward where wormholes emerge from {\it bifurcating}, {\it i.e.} non-extremal black hole horizons \ci{Hayw,Kim}.}

For the AdS case (Fig. \ref{fig:f,bQ>1/2}a), the throat grows and
persists until its mass $M$ vanishes. However, for the cases of
flat (Fig. \ref{fig:f,bQ>1/2}b) and dS (Fig. \ref{fig:f,bQ>1/2}c),
the throat does not grow to an indefinitely large size. There is a
maximum size at a certain mass $\wt{M}$ before M vanishes,
{\it i.e.}, $M^*>\wt{M}>0$. If we further reduce the mass
below $\wt{M}$, only the space-time with a naked singularity is a
possible solution. So, the usual naked singularity solution for
$M<M^*$ may be replaced by the non-singular wormhole geometry
until a certain (non-negative) mass $\wt{M}$ is reached
in flat and dS cases. This may suggest the generalized cosmic
censorship \ci{Cant,Kim} by the existence of a wormhole-like
structure around the origin, which has been thought to be a singular point of black hole solutions where all the usual physical laws {\it could} break down and {\it might} be regarded as an indication of the incompleteness of general relativity.

For the case (ii) of $\beta Q =1/2$, a point-like throat can be generated at the origin contrary to the generation of a finite-size throat for the case (i). But its growing process is the same as the case (i), showing the existence of a maximum size of the throat with the mass $\wt{M}>0$ for flat and dS cases, and $\wt{M}=0$ for AdS case (Fig. \ref{fig:f,bQ=1/2}).

For the case (iii) of $\beta Q <1/2$, the natural wormhole geometry exists only for AdS case (Fig. \ref{fig:f,bQ<1/2}a) and its growing behavior is similar to the case (ii). On the other hand, for flat and dS cases (Fig. \ref{fig:f,bQ<1/2}b, \ref{fig:f,bQ<1/2}c), only the Sch-like type I black hole solution for $M>M_0$ or the naked singularity solution for $0<M<M_0$ can exist.

So far, we have considered geometric aspects in constructing the natural
wormhole. Here, it is important to note that the right-hand side of the Einstein equation depends only on the values of the BI field strength themselves, not on their derivatives. This means that Eq. (\ref{eomforgrav}) is still valid even in the presence of the throat {$r_0$}, where the BI fields do not join smoothly, {\it i.e.}, their (spatial) derivatives are discontinuous (Fig. \ref{fig:E}). This would be valid for any kind of matter field including gauge field if the coupling with the matter and gravity does not contain derivatives. In other words, the same solution of the Einstein equation (\ref{eomforgrav}) without the throat can also be the exact solution for each patch of the wormhole geometry even when considering matter coupled system.

However, the above argument would not hold for the BI field equation
(\ref{eomforBI}), which depends on the derivatives of BI fields.
Equation (\ref{eomforBI}) for the wormhole geometry with a throat can be
modified due to the discontinuities of derivatives of BI fields. To find the possible modification term, let us introduce another coordinate $\eta$ which spans the whole wormhole space-times with a geodesically complete ranges $(-\infty,+\infty)$ by joining two coordinate patches, each one covering $[r_0, +\infty)$ with the throat at $\eta=0$ for $r=r_0$. For example, we may consider the Ellis-type metric {ansatz} for a traversable wormhole, {{\it i.e.}, a throat-type geometry that physical objects can propagate through,}
\begin{\eq}
  ds^2=-N^2(l) dt^2+{dl^2}+r^2({l})\left(d\theta^2+\sin^2\theta d\phi^2\right),
\label{Ellis}
\end{\eq}
where $l$ is the proper length from the throat given by
\beq
l(r)=\pm \int^r_{r_0} {dr}/{\sqrt{f(r)}}.
\eeq
Then, in the region close to the throat $\eta=0$, one may expand the original coordinate $r^{\pm}$ for each patch as
\beq
r^{\pm} (\eta)=r_0 +\left.\left( \f{dr}{d \eta}\right)\right|^{\pm}_{{r_0}} \eta + \f{1}{2}\left.\left( \f{dr^2}{d \eta^2}\right)\right|^{\pm}_{{r_0}} \eta^2+\cdots.
\label{r:eta}
\eeq
Here, we note that $ ({dr}/{d \eta})|^+_{{r_0}}\geq 0,
~({dr}/{d \eta})|^-_{{r_0}} \leq 0$, and
$({dr^2}/{d \eta^2})|^\pm_{{r_0}}>0$
in order that the minimum radius $r_0$ exists.

In general, there are two options for the choice of $\eta$ coordinate. \\

\begin{figure}
\includegraphics[width=10cm,keepaspectratio]{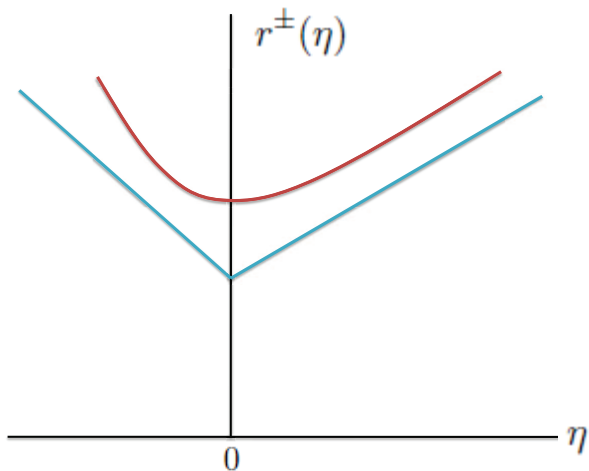}
\caption{The plots of possible $r(\eta)$ functions.   }
\label{fig:r:eta}
\end{figure}

(a) First, if we consider the regular coordinate transformation $r^{\pm}(\eta)$ of (\ref{r:eta}) without singularity, we need to consider $({dr}/{d \eta})|^+_{{r_0}}=({dr}/{d \eta})|^-_{{r_0}}=0$ but its inverse transformation $\eta(r^{\pm})$ becomes singular at the throat, $({d\eta}/{d r})|^\pm_{{r_0}} \ra \pm \infty$ (top curve in Fig. \ref{fig:r:eta}) \ci{Elli,Viss}. \\

(b) Second, if we consider the case of $({dr}/{d \eta})|^\pm_{{r_0}}\neq0$ (bottom curve in Fig. \ref{fig:r:eta}), we have a singularity -- delta function singularity due to discontinuity of $({dr}/{d \eta})^\pm_0$-- in the third term of (\ref{r:eta}) so that the coordinate transformation $r^{\pm} (\eta)$ becomes singular at the throat.\\

The second option was not considered before \ci{Elli,Viss} but it seems that both options are possible on the general ground. For example, if we consider the derivative of the metric tensor $g^{\pm}_{\m\n}(r)$ with respect to the proper length $l$ in the \Sch coordinate (\ref{wormhole})
\beq
\f{dg^{\pm}_{\m\n}(r)}{dl}=\left(\f{dr^{\pm}}{dl} \right) \f{dg^{\pm}_{\m\n}(r)}{dr^{\pm}}=\pm \sqrt{f^{\pm}(r)} ~ \f{dg^{\pm}_{\m\n}(r)}{dr^{\pm}},
\label{der_l}
\eeq
one can join the metric smoothly, ${dg^{+}_{\m\n}(r)}/{dl}|_{r_0}={dg^{-}_{\m\n}(r)}/{dl}|_{r_0}=0$, by considering either $f^{\pm}(r_0)=0$ (Morris-Thorne approach \ci{Elli}) or ${dg^{+}_{\m\n}}/{dr}|_{r_0}={dg^{-}_{\m\n}}/{dr}|_{r_0}=0$
(new approach in this paper \ci{Cant,Kim}).

\begin{figure}
\includegraphics[width=7cm,keepaspectratio]{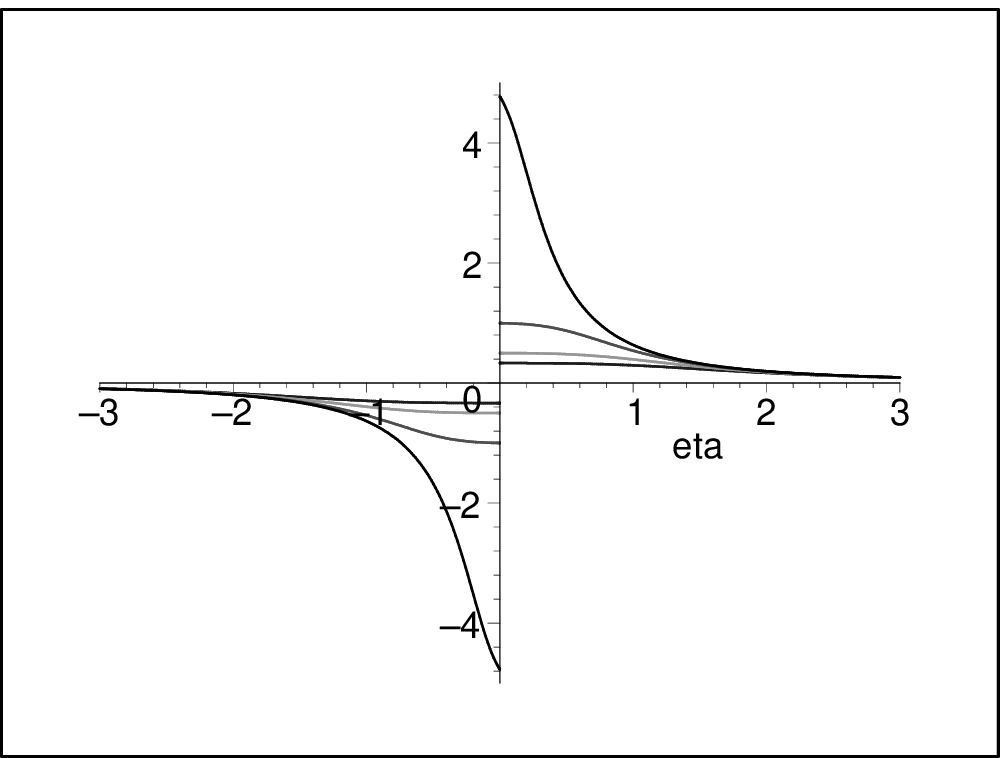}
\qquad
\includegraphics[width=8cm,keepaspectratio]{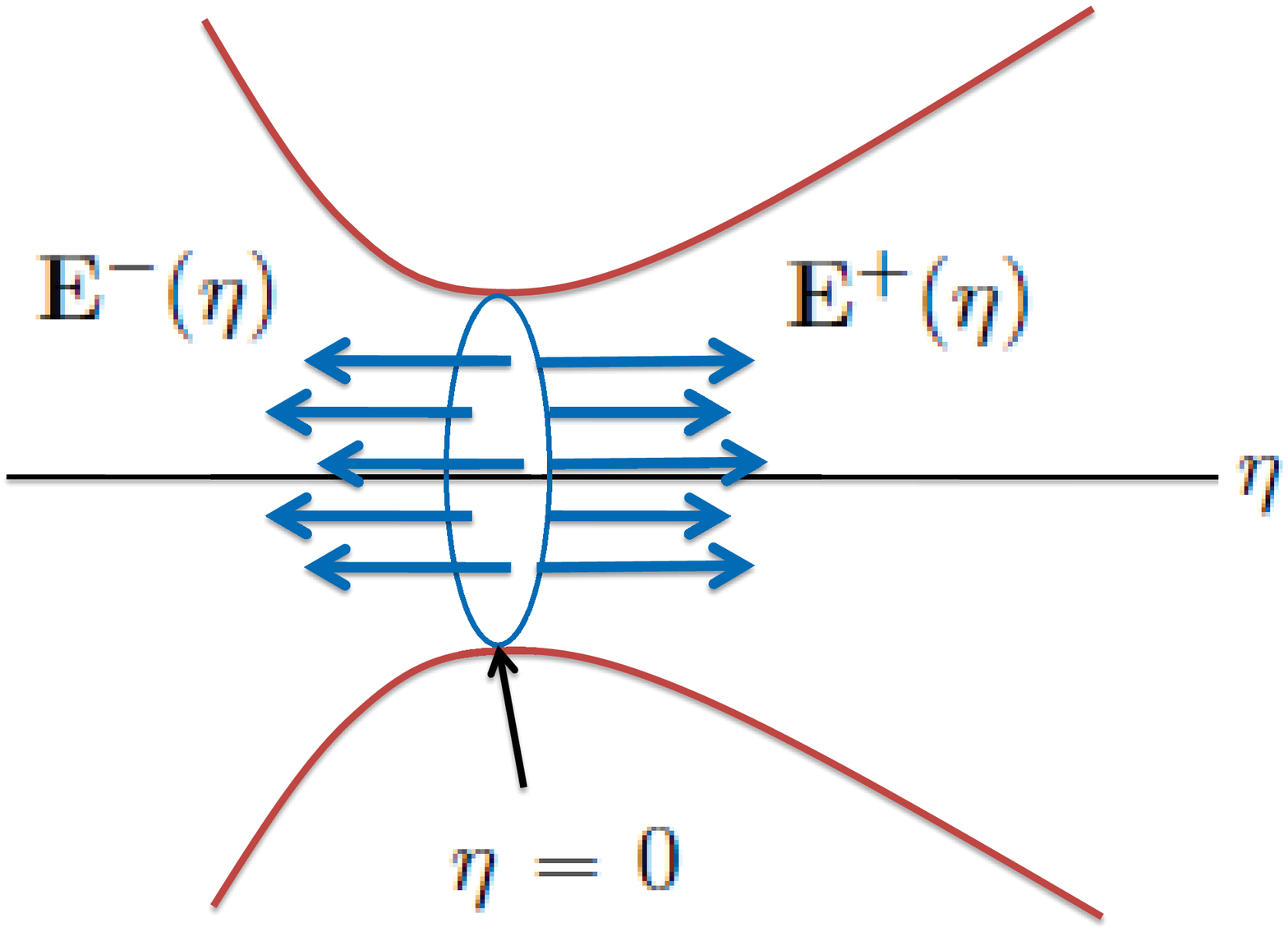}
\caption{The plots of $E(\eta)$ and ${{\bf E}^{\pm}(\eta)}=E(\eta) \hat{\eta} $ functions.   }
\label{fig:E:eta}
\end{figure}

But, independently of the choice of the two options, we may generally
obtain the radial component of the BI electric field and its derivative,
\beq
E(r)&=&E^+(r) \ep(\eta)+E^-(r) \ep(-\eta), \no \\
\f{d E(r)}{dr}&=&\f{dE^+(r)}{dr} \ep(\eta)+\f{d E^-(r)}{dr} \ep(-\eta)+E^+(r_0)
\left[\left.\left(\f{d \eta}{d r} \right)\right|^+_{r_0}  +\left.\left(\f{d \eta}{d r} \right)\right|^-_{r_0} \right]\de(\eta),
\eeq
where we have used the continuity of the field $E^+(r_0)=-E^-(r_0)$ and
$d\ep(\pm \eta)/d \eta=\pm \de(\eta)$ with the step function
$\ep(\eta)$ (Fig. \ref{fig:E:eta}).

In general, singularities may enter into the BI field equation when it involves the radial derivative {$\f{d}{dr}$} for the wormhole geometry with a throat, regardless of the choice of $\eta$ coordinate unless we {\it fine-tune} the coordinate $\eta$ \ci{Elli,Viss} so that
\beq
\left.\left(\f{d \eta}{d r} \right)\right|^+_{r_0}  =-\left.\left(\f{d \eta}{d r} \right)\right|^-_{r_0} .
\label{tune}
\eeq
Note that the condition (\ref{tune}) has been {\it implicitly} assumed in the
Morris-Thorne construction for the first option, though
$(d \eta/dr)^{\pm}_{r_0}$ becomes infinity \ci{Elli}. Now, in order to
understand the condition (\ref{tune}) in our construction for the second
option, let us consider the second derivatives of the metric tensor
$g^{\pm}_{\m\n}(r)$ with respect to $\eta$
\beq
\f{d^2g^{\pm}_{\m\n}(r)}{d\eta^2}=
\left(\f{dr^{\pm}}{d\eta} \right)^2 \f{d^2g^{\pm}_{\m\n}(r)}{d{r^{\pm}}^2}
+\left(\f{d^2r^{\pm}}{d\eta^2} \right) \f{dg^{\pm}_{\m\n}(r)}{dr^{\pm}}.
\label{der_r_r}
\eeq
Then, the continuity of the second derivatives with respect to $\eta$ and $r$ at the throat (note that the second term in (\ref{der_r_r}) vanishes in our case),
\beq
\left.\left(\f{d^2g^{+}_{\m\n}}{d\eta^2}\right)\right|_{r_0}
=\left.\left(\f{d^2g^{-}_{\m\n}}{d\eta^2}\right)\right|_{r_0},~
\left.\left(\f{d^2g^{+}_{\m\n}}{d{r^+}^2}\right)\right|_{r_0}
=\left.\left(\f{d^2g^{-}_{\m\n}}{d{r^-}^2}\right)\right|_{r_0}
\eeq
requires the condition (\ref{tune}) naturally. Note that, in the usual case of the first option, the first term in (\ref{der_r_r}) vanishes and one can obtain another condition
$\left(\f{d^2r^{+}}{d\eta^2}\right)_{r_0}
=\left(\f{d^2r^{-}}{d\eta^2}\right)_{r_0}$
from the continuity of the first derivative with respect to $r$,
$\left(\f{dg^{+}_{\m\n}}{dr^+}\right)_{r_0}
=\left(\f{dg^{-}_{\m\n}}{dr^-}\right)_{r_0}$, which need not to be vanished \ci{Elli,Viss}. On the other hand, in our case of the second option, the condition (\ref{tune}) is also necessary in order to have the continuity of the metric tensor and its derivatives in a ``coordinate independent'' way. Here, it is important to note that there is no constraint on $\left(\f{d^2r^{\pm}}{d\eta^2}\right)$ which is crucial for the discussion of the energy condition in Sec. IV.

Another important consequence of the condition (\ref{tune}) is the consistency of the field equation (\ref{eomforBI}) with the Einstein equation (\ref{eomforgrav}). In the usual equations of motion, we need the Bianchi identity for the Riemann tensors and the continuity equation for the (matter's) energy-momentum tensors
\beq
\nabla_{\mu} T^{\m \n}= 0 .
\label{flow_eq}
\eeq
But, instead of considering the continuity equation (\ref{flow_eq}), there is an easier way to check it. It is to consider the general relation before implementing (\ref{flow_eq}),
\beq
p_{\theta}=p_{\phi}=-{\rho}-\f{r}{2} \f{d}{dr} \rho-\f{r}{2}\nabla_{\mu} T^{\m}_r ,
\label{p_rho_general}
\eeq
for the energy-momentum tensor
$ T^{\m}_{ \n}{(r)}= {\rm{diag}}(-\rho,p_r,p_{\theta},p_{\phi})$ with
\beq
\rho&=&-p_r=-4 \beta^2 \left(1-\f{1}{\sqrt{1-E^2/\beta^2}} \right),\no \\
~p_{\theta}&=&p_{\phi}=4 \beta^2 \left(1-{\sqrt{1-E^2/\beta^2}} \right).
\label{p_rho_sol}
\eeq
Note that, when the usual continuity equation (\ref{flow_eq}) is used, we have the conventional relation (see Ref. \ci{Nico}, for example),
\beq
p_{\theta}=p_{\phi}=-{\rho}-\f{r}{2} \f{d}{dr} \rho.
\label{p_rho}
\eeq
In other words, if the usual relation (\ref{p_rho}) is violated, the
continuity equation (\ref{flow_eq}) is also violated exactly at the same
place where (\ref{p_rho}) fails. Actually, in our case, (\ref{p_rho}) fails
at the throat if the condition (\ref{tune}) is not considered ! From
Eq. (\ref{p_rho_sol}), $(p_\theta,p_\phi)$ of (\ref{p_rho}) depend only on
the field $E(r)$, not on its derivatives $dE(r)/dr$ so that they are not
affected by the presence of the throat. However the right hand side of
(\ref{p_rho}) depends on $dE(r)/dr$, as well as $E(r)$, which introduces the
additional delta-function term at the throat. This means that, in order to
have a ``consistent'' equation (\ref{p_rho}), we need a counter term on the
right hand side to cancel the singular term at the throat implying the violation of the continuity equation
\beq
\nabla_{\mu} T^{\m \n}\sim \de(\eta) .
\label{flow_eq_viol}
\eeq
Because the Einstein equation (\ref{eomforgrav}) is not affected by the existence of the natural wormhole, this violation of the continuity equation means that the Bianchi identity for the curvature tensor fails also at the throat,
\beq
\nabla_{\mu} G^{\m \n}\sim \de(\eta) ,
\label{Bianchi_viol}
\eeq
with the Einstein tensor $G_{\m\n}=R_{\m \n}-\f{1}{2} R g_{\m \n}.$ In electrodynamics, the Bianchi identity $\pa_{\m}\tilde{F}^{\m \n}=0$, with the dual field strength $\tilde{F}^{\m \n}=\f{1}{2} \ep^{\mu \n \sigma \rho} F_{\sigma \rho},$ is violated at the location of the magnetic monopole. In our case, this electromagnetic Bianchi identity works as one can easily check from the solution (\ref{Esol}) since $\tilde{F}^{\m \n}=0$. But we may have the violation of the gravitational Bianchi identity at the throat if we do not require the condition (\ref{tune}).\footnote{This will apply also to ``strings and other distributional sources" in GR \ci{Gero:78}. We thank Jennie Traschen for pointing out this.} This can be considered as another support of the condition (\ref{tune}) in our new approach.

\section{Energy conditions}

In the usual set-up of wormholes, exotic matters violating the energy conditions are essential to sustain the throats.\footnote{This result assumes topologically trivial and static matters. So, there could still exit some rooms for avoiding the no-go result by relaxing those assumptions \ci{Ayon}.} Even when we look for
wormhole solutions without the explicit introduction of exotic matters,
we need some effective matter terms in the Einstein equation which play the
role of exotic matters. Wormhole solutions in higher-derivative gravities is one example. In that case, higher-derivative terms act as the effective energy-momentum tensor which can violate the energy conditions generally. However, this is not the case in our construction of natural wormholes and all the energy conditions can be normal.

\begin{figure}
\includegraphics[width=10cm,keepaspectratio]{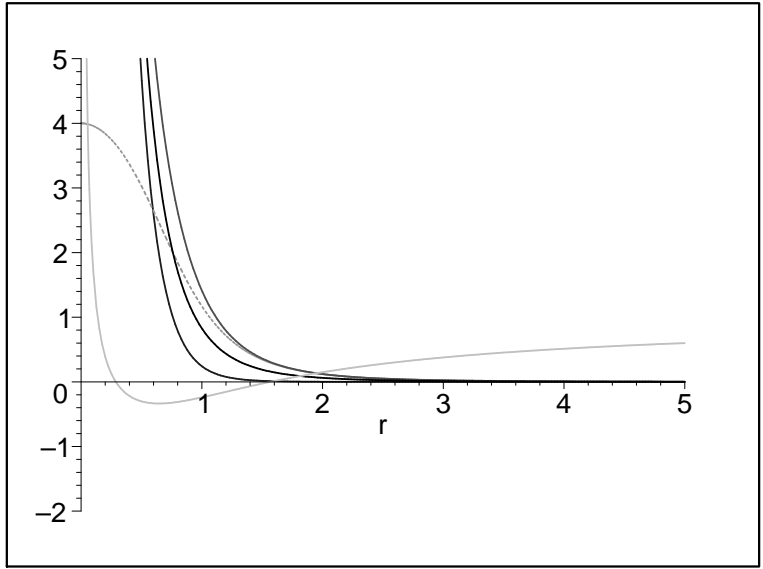}
\caption{The plots of $\rho-p_\theta$, $\rho$, $\rho+p_\theta$, (three thick curves, left to right), $\rho+\sum_i p_{i}$ (thin dotted curve), and $f(r)$ (thin solid curve) for $\beta=1,~ Q=1$.}
\label{fig:energy_cond}
\end{figure}

To see this, let us consider
\beq
\rho+p_{\theta}&=&\rho+p_{\phi}
=4 \beta^2 \left(\f{1}{\sqrt{1-E^2/\beta^2}} -{\sqrt{1-E^2/\beta^2}} \right),
\label{rho+P} \\
\rho-p_{\theta}&=&4 \beta^2 \left(\f{1}{\sqrt{1-E^2/\beta^2}} +{\sqrt{1-E^2/\beta^2}}-2 \right),
\label{rho-P} \\
\rho+\sum_i p_{i}&=&2 p_{\theta}=
4 \beta^2 \left(1-{\sqrt{1-E^2/\beta^2}}\right),
\label{rho+Sum_P}
\eeq
where we have used $\rho+p_r=0$ from (\ref{p_rho_sol}). Then, it is easy to see that the quantities in (\ref{rho+P})$\sim$(\ref{rho+Sum_P}), as well as $\rho$ in (\ref{p_rho_sol}), are all non-negative for the whole range of $E(r) \leq \beta$. In Fig. \ref{fig:energy_cond}, we have plotted the case for two horizons in the RN-like type II black hole but the result is the same as the Sch-like type I case. This shows that the strong energy condition (SEC: $\rho+p_i \geq 0, ~\rho+\sum_i p_{i} \geq 0$), which includes the null energy condition (NEC: $\rho+p_i \geq 0$), is satisfied as well as the weak energy condition (WEC: $\rho \geq 0, ~\rho+p_{i} \geq 0$) and the dominant energy condition (DEC: $\rho \geq |p_i|$ or equivalently $\rho \pm p_i \geq 0$).

In order to understand the physical origin of this result, which is
unusual in the conventional wormhole physics, let us consider the
usual Morris-Thorne ansatz instead of (\ref{wormhole}) {\ci{Elli,Viss}},
\begin{\eq}
  ds^2=-e^{2 \phi_{\pm}(r)}dt^2+\frac{dr^2}{1-b_{\pm}(r)/r}+r^2
\left(d\theta^2+\sin^2\theta d\phi^2\right).
\label{MT}
\end{\eq}
Then, the energy density and pressures are given by
\beq
\rho &=& {2}
\left[ \f{1}{r^2}\f{db}{dr}-\Lambda\right],
\label{rho_MT}\\
p_r &=& {2}
\left[ -\f{b}{r^3}+2\left(1-\f{b}{r}\right)
{\f{1}{r}\f{d \phi}{dr}}+\Lambda\right],
\label{p_r_MT} \\
p_\theta &=&p_\phi={2}
\left\{ \left(1-\f{b}{r} \right)\left[\f{d^2\phi}{dr^2}+\f{d\phi}{dr} \left(\f{d\phi}{dr}+\f{1}{r} \right) \right] -\f{1}{2r^2}\left(r \f{db}{dr} -b \right) \left(\f{d\phi}{dr}+\f{1}{r}\right)+\Lambda\right\},
\label{p_theta_MT}
\eeq
from the Einstein equation, $G_{\m \n}+\La g_{\m \n}=T_{\m \n}{/2}$.

An important combination of the quantities in
(\ref{rho_MT}) $\sim$ (\ref{p_theta_MT}), which is crucial for the violation of energy condition in the usual approach \ci{Elli,Viss}, is
\beq
\rho+p_r=-\f{{2} e^{2 \phi}}{r} \f{d}{dr}\left[\f{e^{-2 \phi}}{r} \left(1-\f{b}{r} \right) \right].
\eeq
In the usual approach of the first option, the quantity in the bracket vanishes at the throat $r=r_0$ since, when we consider the proper length $l$, for example, $({dr}/{d l})|_{r_0}=\pm \sqrt{1-b(r_0)/r_0}=0$, {\it i.e.}, $b(r_0)=r_0$. But, in order that $b(r)<r$ for $r>r_0$ so that $r$ is the space-like coordinate for an observer outside the throat, the quantity in the bracket needs to be positive, {\it i.e.}, it has ``positive" derivative \footnote{At the throat $r=r_0$, this is equivalent to $(d^2 r/d l^2)|_{r_0}=\f{1}{2} \f{d}{dr}(1-b/r)|_{r_0} >0$, which is known as the ``flaring-out" condition \ci{Elli}.}
\beq
\f{d}{dr}\left[\f{e^{-2 \phi}}{r} \left(1-\f{b}{r} \right) \right] >0,
\eeq
which reduces to
\beq
\rho+p_r <0,
\eeq
so that all the energy conditions are violated. However, this is not the case in our approach of the second option, where the quantity in the bracket does not vanish at the throat for the traversable wormhole. The radial derivative of the quantity in the bracket, which is {a} positive quantity for $r>r_0$, can have any values. Actually, in our example, we have $\rho+p_r =0$. One can also obtain similar results for other combinations of the energy and pressures as in Fig. \ref{fig:energy_cond}. This shows that the exoticity of the energy-momentum tensor for matters or effective matters from the modified gravities is {\it not} essential in the new approach.

\section{Discussion}

We have studied a new approach to construct a natural wormhole geometry in
EBI gravity without introducing additional exotic matters at the throat.
BI field equations are not modified at the throat for {\it coordinate
independent} conditions of continuous metric tensor and its derivatives,
even though BI fields have discontinuities in their derivatives generally.
If we do not require the newly introduced conditions, the modification term exists and it produces the violation of the gravitational Bianchi identity.
This supports the necessity of the new conditions in our construction.
Remarkably, there is no violation of energy conditions and the
energy-momentum tensors are normal. It is shown that this is the crucial
effect of the new approach and the exoticity of the energy-momentum tensors for the matters is not essential for natural wormholes. It would be a challenging problem to see whether the non-exoticity in the energy-momentum tensors is a sign for the stability of wormholes or not. Several further remarks are in order.

First, the new approach for wormholes has been first studied in
Ho\v{r}ava gravity \ci{Cant}, which has been proposed as a renormalizable
quantum gravity model, where the effective energy-momentum tensors from
the higher spatial derivative contributions violate the usual energy
conditions in general. There, it has been argued that the Ho\v{r}ava gravity wormhole is the result of microscopic wormholes created by negative
energy quanta which have entered the black hole horizon in Hawking radiation process
\footnote{Hawking radiation process involves virtual pairs of particles
near the event horizon, one of the pair enters into the black hole while the other escapes. The outgoing particle can be observed as a real particle with a positive energy with respect to an observer at infinity. Then, the ingoing particle must have a {\it negative} energy in order that the energy is conserved \ci{Hawk}. This implies that the negative energy particle that falls into black hole can play the role of exotic matter for wormhole formation \ci{Kim}.}. In other words, the existence of wormholes may reflect the quantum gravity effect of black holes. However, we have found that the natural wormhole construction is still valid in the classically charged black holes in GR and the usual energy conditions can be satisfied too. This indicates that the formation mechanism for the natural charged wormholes has the classical origin in contrast to {\it vacuum} wormholes in Ho\v{r}ava gravity and other higher curvature gravities. So, there are two possible origins for wormholes, one for the classical and the other for the quantum mechanical ones, and both effects need to be considered generally. Actually, in our wormholes for EBI gravity, there is the upper bound for the electric field which may reflect the quantum mechanical pair creation of photons \footnote{In the string theory context, where BI type action is considered as its low energy effective action, this corresponds to the pair creation of open strings.} at short distances so that both the classical and quantum effects are involved in the natural wormholes.

Second, we note that the new approach may have some difficulties when extending to higher derivative gravity theories due to possible singularities at the throat. For example, when third-order spatial derivatives appear in the Einstein equation, we need to consider
\beq
\f{d^3g^{\pm}_{\m\n}(r)}{dl^3}=
\left(\f{dr^{\pm}}{dl} \right)^3 \f{d^3g^{\pm}_{\m\n}(r)}{d{r^\pm}^3}
+3 \left(\f{d^2r^{\pm}}{dl^2} \right) \left(\f{dr^{\pm}}{dl} \right) \f{d^2g^{\pm}_{\m\n}(r)}{d{r^\pm}^2}
+\left(\f{d^3r^{\pm}}{dl^3} \right) \f{dg^{\pm}_{\m\n}(r)}{dr^{\pm}}.
\label{der_r_r_r}
\eeq
In the usual Morris-Thorne approach, only the third term survives at
the horizon due to $(dr^{\pm}/dl)|_{r_0}=0$ and it remains finite from the continuity condition $(d^2r^{+}/dl^2)|_{r_0}=(d^2r^{-}/dl^2)|_{r_0}$. Similarly, one may generalize this argument to arbitrarily higher-order derivatives in the Einstein equation. On the other hand, in the new approach the third term vanishes but the first two terms survive at the throat due to
$(dg^{\pm}_{\m \n}/dr^{\pm})|_{r_0}=0$. However, the second term can
cause a singularity due to the discontinuity of $(dr^{\pm}/dl)|_{r_0}$
unless we consider the case of $(d^2g^{\pm}_{\m \n}/d{r^\pm}^2)|_{r_0}=0$,
which does not seem to be quite generic. This might imply that the new
approach is not complete to describe all the generic higher-derivative
gravity theories. Or this might imply that the natural wormholes are
unstable for the case where the higher derivatives are involved like
rotating black holes in  Ho\v{r}ava gravity or other higher curvature
gravities, as has been pointed out by one of us recently \ci{Kim}.\\

Note added: After finishing this paper, a related paper \ci{Li} appeared whose analysis of black hole in four dimensions is partly overlapping with ours.

\section*{Acknowledgments}

JYK appreciates {\it National Center for Theoretical Sciences} (Hsinchu) and
C. Q. Geng for hospitality during the preparation of this work.
MIP appreciates {\it The Interdisciplinary Center for Theoretical Study}
(Hefei) where this work was completed and JianXin Lu, Li-Ming Cao, and
other faculty members for hospitality during his visit. MIP was supported by Basic Science Research Program through the National Research
Foundation of Korea (NRF) funded by the Ministry
of Education, Science and Technology {(2016R1A2B401304)}.

\newcommand{\J}[4]{#1 {\bf #2} #3 (#4)}
\newcommand{\andJ}[3]{{\bf #1} (#2) #3}
\newcommand{\AP}{Ann. Phys. (N.Y.)}
\newcommand{\MPL}{Mod. Phys. Lett.}
\newcommand{\NP}{Nucl. Phys.}
\newcommand{\PL}{Phys. Lett.}
\newcommand{\PR}{Phys. Rev. D}
\newcommand{\PRL}{Phys. Rev. Lett.}
\newcommand{\PTP}{Prog. Theor. Phys.}
\newcommand{\hep}[1]{ hep-th/{#1}}
\newcommand{\hepp}[1]{ hep-ph/{#1}}
\newcommand{\hepg}[1]{ gr-qc/{#1}}
\newcommand{\bi}{ \bibitem}

\end{document}